\documentclass[a4paper]{jpconf}
\usepackage{graphicx}
\begin{document}
\title{Spherical gravitational collapse in 5D Einstein-Gauss-Bonnet gravity}
\author{Sushant G. Ghosh$^1$, Sanjay Jhingan$^1$, D. W. Deshkar$^2$}
\address{$^1$Center for Theoretical Physics, Jamia Millia Islamia,
New Delhi 110025, India}
\address{$^{2}$ S.S.E.S. Amravati's Science College,
Congress Nagar,  Nagpur-440 012, India}

\ead{sgghosh@gmail.com,sanjay.jhingan@gmail.com,dwdeshkar@gmail.com}

\begin{abstract}
We consider a spherical gravitational collapse of inhomogeneous dust
(and null dust) in Einstein gravity with the Gauss-Bonnet (GB)
combination of quadratic curvature terms.  It turns out that the
presence of the coupling constant of the GB terms $\alpha
> 0$ completely changes the causal structure of the singularities
from the analogous general relativistic case. The gravitational
collapse of inhomogeneous dust in the five-dimensional GB extended
Einstein equations leads to formation of a massive, but weak,
timelike singularity which is forbidden in general relativity. The
apparent horizons of two different collapsing solutions  show
interesting mathematical similarity.
\end{abstract}

\section{Introduction}
The Lemaitre-Tolman-Bondi (LTB) \cite{tb} and Vaidya \cite{pc}
solutions  have been extensively used not only to study the
formation of naked singularities and black holes in spherical
collapse, but in cosmology as well. It is well known that these two
solutions admit both naked and covered singularities depending upon
the choice of initial data and there is a smooth transition from one
phase to the other \cite{r1}. These results have led to strong
evidence against the cosmic censorship conjecture (CCC) \cite{rp},
which asserts that \emph{there can be no singularity visible from
future null infinity}. In recent years a renewed interest has grown
in higher order gravity, which involves higher derivative curvature
terms, and amongst the most extensively studied theory is the
so-called Einstein-Gauss-Bonnet (EGB) gravity. It appears naturally
in the low-energy effective action of heterotic string theory
\cite{Gross}. Boulware and Deser \cite{bd} found exact black hole
solutions in D $(\geq 5)$-dimensional gravitational theories with a
Gauss-Bonnet (GB) term  modifying the usual Einstein-Hilbert action.
Recently, Maeda \cite{maeda} considered the spherically symmetric
gravitational collapse of a inhomogeneous dust with the D $(\geq
5)$-dimensional action including the GB term. He investigated its
effects on the final fate of gravitational collapse without finding
the explicit form of the solution.

In this paper, we consider the 5-dimensional (5D)  action with the
GB terms for gravity and give an {\em exact model} of the
gravitational collapse of a inhomogeneous dust including the second
order perturbative effects of quantum gravity. We also consider the
effects of GB terms on the structure of apparent horizons in dust
solutions as well as in null dust solutions. The nature of
singularities of such a spacetime in terms of its being hidden
within a black hole, or whether it would be visible to outside
observers, and the consequence of GB term on 5D-LTB collapse are
analyzed.

\section{Basic equations in EGB}
The gravitational part of the   5D action that we consider is
\cite{gd,hm}:
\begin{equation}
\label{action} S=\int
d^5x\sqrt{-g}\biggl[\frac{1}{2\kappa_5^2}(R-2\Lambda+\alpha{L}_{GB})
\biggr]+S_{\rm matter},
\end{equation}
where $R$ and $\Lambda$ are the 5D Ricci scalar and the cosmological
constant respectively. $\kappa_5\equiv\sqrt{8\pi G_5}$, where $G_5$
is the 5D gravitational constant. The GB Lagrangian is the
combination of the Ricci scalar, Ricci tensor $R_{ab}$, and Riemann
tensor $R^a_{~~b\rho\sigma}$ as $
{L}_{GB}=R^2-4R_{ab}R^{ab}+R_{ab\rho\sigma}R^{ab\rho\sigma}. $ In
the 4-dimensional spacetime, the GB terms do not contribute to the
field equations. The $\alpha$ is the coupling constant of the GB
terms. From the action (\ref{action}) we derive the following field
equations:
\begin{eqnarray}
& & {G}_{a b} - \alpha {H}_{a b} =  {T}_{ab}, \mbox{with
 }\; {G}_{ab} = R_{ab}-{1 \over 2}g_{ab}R,\; \\ \nonumber  & &
{H}_{ab}=2\Bigl[RR_{ab}-2R_{a \alpha}R^{\alpha}_{b}-2R^{\alpha
\beta}R_{a \alpha b\beta}
 +R_{a}^{~\alpha\beta\gamma}R_{b \alpha\beta\gamma}\Bigr]
-{1\over 2}g_{ab}{L}_{GB},
\end{eqnarray}
where $G_{ab}$ is the Einstein tensor and  $H_{ab}$ is the Lanczos
tensor.
\section{Null dust collapse} The energy-momentum tensor of a null
fluid is ${T}_{ab}=\psi(v,r) l_{a}l_{b} $,  where $\psi(v,r)$ is the
non-zero energy density and $l_a$ is a null vector such that $l_{a}
= \delta_a^0,\; \mbox{with}\; l_{a}l^{a} = 0. $ Expressed in terms
of Eddington advanced time coordinate (ingoing coordinate) $v$, the
metric of general spherically symmetric spacetime \cite{gd,hm}:
\begin{equation}
ds^2 = - A(v,r)^2 f(v,r)\;  dv^2
 +  2 A(v,r)\; dv\; dr + r^2 d \Omega^2_3, \label{eq:megb}
\end{equation}
$ d\Omega^2_3 = d \theta^2+ \sin^2 \theta d \phi^2 + \sin^2 \theta
 \sin^2\phi\; d\psi^2$. Here $A(v,r)$ is an arbitrary function.
 It is the field equation $G^0_1 = 0$ that
leads to $ A(v,r) = g(v)$. This could be absorbed by writing $d
\tilde{v} = g(v) dv$. Then, $f(v,r)$ is obtained by solving field
equation to give the general solution as
\begin{equation}
f(v,r) = 1 + \frac{r^2}{4 \alpha} \left[ 1 \pm \sqrt{1 + \frac{8
\alpha m(v)}{r^4}}\right]; \; \psi(v,r)=\frac{3}{2r^{3}}{\dot m}(v),
\label{s-feqgb}
\end{equation}
where $m(v)$ is an arbitrary function of $v$. There are two families
of solutions which correspond to the sign in front of the square
root in Eq.~(\ref{s-feqgb}), which has the minus (plus) sign
respectively for the minus- (plus+) branch solutions.
 The energy density of the null fluid as is given by $\psi(v,r)$ for both branches, where
the dot denotes the derivative with respect to $v$.  In the general
relativistic limit ${ \alpha} \to 0$, the minus-branch solution
reduces to the 5D Vaidya solution in Einstein gravity. In the static
case ${\dot m}=0$, this solution reduces to the solution which was
independently discovered by Boulware and Deser~\cite{bd} and
Wheeler~\cite{jtw}. It has been shown that a timelike naked
singularity is formed from gravitational collapse of null dust in
EGB, which does not appear in the general relativistic case
\cite{hm}.

 The apparent horizon (AH) is the outermost marginally trapped surface
for the outgoing photons.    The AH are defined as surface such that
$\Theta \simeq 0$ which implies that $f=0$.  It is clear that AH
\cite{gd} is the solution of
\begin{equation}\label{ahnf}
\left[1 + \frac{r^2}{4\alpha}\left[1 - \sqrt{1 + 8\alpha
\left(\frac{m(v)}{r^4}\right)}\right]\right] = 0,\; \mbox{implies}\;
r_{AH} = \sqrt{m(v) - 2\alpha}.
\end{equation}
In the relativistic limit $\alpha \rightarrow 0$ then $r_{AH}
\rightarrow \sqrt{m(v)}$. One sees that $g_{vv} = 0$ implies that $r
= \sqrt{m(v) - 2 \alpha}$ is timelike surface \cite{gd}.

\section{Inhomogeneous dust collapse}
The solution we seek is  collapse of a spherical dust in 5D-EGB. The
energy-momentum tensor for dust is $T_{ab} =  \epsilon(t,r)
\delta_{a}^t \delta_{b}^t$ where $u_a = \delta_t^a$ is the 5D
velocity. The metric for the 5D case, in comoving coordinates, is
\cite{sjsg,gb,gab}:
\begin{equation}\label{metric}
ds^2 = -dt^2 + A(t,r)^2 dr^2 + R(t,r)^2 d\Omega_3^2.
\end{equation}
 The coordinate $r$ is the comoving radial coordinate, $t$ is the proper time of freely falling
shells, $R$ is a function of $t$ and $r$ with $R \geq 0$, and $A$ is
also a function of $t$ and $r$. The master equation of the system is
given by
\begin{equation}
\dot{R}^2 \left[1 - 4 \alpha\frac{W^2-1}{R^2} \right] = (W^2 - 1) +
\frac{F}{R^2} - 2 \alpha \frac{\dot{R}^4}{R^2}.\label{eq:fe}
\end{equation}
Here $F = F(r)$ is an arbitrary function of $r$ and is referred to
as mass function.  Indeed the energy density, $\epsilon =
{3F'}/{2R^3 R'}$, must be non-negative.  It is easy to see that as
$\alpha \rightarrow 0$ the master solution (\ref{eq:fe}) of the
system reduces to the corresponding 5D-LTB solution in \cite{gab}.
In the present discussion, we are concerned with gravitational
collapse, which requires $\dot{R}(t,r) < 0$. Eq.~(\ref{eq:fe}), for
$W(r)=1$, can be integrated  to \cite{sjsg}
\begin{eqnarray}\label{solution}
{t_\varsigma(r)-t}  = {\frac{\sqrt{\alpha}}{2\sqrt{2}}} \tan^{-1}
\left[\frac{3 R^2 -\sqrt{R^4+8\alpha F}}{2\sqrt{2} R [\sqrt{R^4 +
8\alpha F} -R^2]^{1/2}} \right]  +\sqrt{\frac{\alpha
R^2}{\sqrt{R^4+8\alpha F}-R^2}}\;, \quad
\end{eqnarray}
where $t_\varsigma(r)$ is an arbitrary function of integration. The
mass function $F$ can be related with initial data (density) at the
scaling surface, $t=0$ $(R=r)$, reduces to form $F(r) =
\frac{2}{3}\int_0^{r} \epsilon(0,r) r^3 dr , $ which completely
specifies the mass function in terms of the initial density profile.
The function $F$ must be positive, because $F < 0 $ implies the
existence of negative mass. This can be seen from the mass function
$m(t,r)$ \cite{gab}, which in the 5D-LTB-EGB case is given by
\begin{eqnarray}
m(t,r) & = & R^2 \left(1 - g^{ab} R_{,a} R_{,b} \right)   = R^2
\left(1 - \frac{{R'}^2}{A^2} + \dot{R}^2 \right). \label{eq:m2}
\end{eqnarray}
Using Eq.~(\ref{eq:fe}) into Eq.~(\ref{eq:m2}), we get $m(t,r) =
F(r) - 2 \alpha \dot{R}^4.$ It may be noted that one can also
calculate mass using the formula proposed by Maeda \cite{maeda} for
the generalized mass function in the EGB. The mass function $F(r) =
m(t,r) + 2 \alpha \dot{R}^4$, is equivalent, up to a constant
factor, to the generalized mass function in EGB \cite{maeda}.

The main advantage of working with the AH is that it is local in
time and can be located at a given spacelike hypersurface.
Considering Eq. (\ref{eq:fe}), the AH condition \cite{sjsg}  becomes
\begin{equation}\label{app-cond}
R(t_{AH}(r),r) =  \sqrt{F(r) - 2 \alpha}.
\end{equation}
It is clear that the presence of  the coupling constant of the GB
terms $\alpha$ produces a change in the location of these horizons.
In the relativistic limit, $\alpha \rightarrow 0$, $R_{AH}
\rightarrow \sqrt{F(r)}$ \cite{gb}. For nonzero $\alpha$, the
structure of the AH is non-trivial. Interestingly, the theory
demands $\alpha$ to be a positive number which forbids AH from
reaching the center, thereby making the singularity massive and
eternally visible, which is forbidden in the corresponding general
relativistic scenario. In general relativity,  noncentral
singularity is always covered \cite{dc} (see also \cite{cjjs}).
However, in the presence of the GB term we find that even the
noncentral singularity is naked, in spite of being massive
([$F(r>0)]
> 0$). Further, Eq.~(\ref{app-cond}) has a mathematical similarity
for the AH given by Eq.~(\ref{ahnf}) in null fluid collapse
\cite{gd}.

Equation~(\ref{app-cond}) implicitly defines a curve $t_{ah}(r)$ and
represents the AH, i.e. the time at which the shell gets trapped. To
further analyze the horizon curve, We use singularity condition in
Eq.~(\ref{solution}) to get
\begin{eqnarray}\label{new_dynamics}
{t_c(r)-t} &=& \frac{\pi\sqrt{\alpha}}{4\sqrt{2}} +
\sqrt{\frac{\alpha R^2}{\sqrt{R^4+8\alpha F}-R^2}}
+{\frac{\sqrt{\alpha}}{2\sqrt{2}}} \tan^{-1} \left[\frac{3 R^2
-\sqrt{R^4+8\alpha F}}{2\sqrt{2} R [\sqrt{R^4 + 8\alpha F}
-R^2]^{1/2}} \right].
\end{eqnarray}
Then, the AH condition (\ref{app-cond}) reduces
Eq.~(\ref{new_dynamics}) to form
\begin{eqnarray}\label{sin-app}
t_c(r) - t_{AH}(r) = \frac{\pi \sqrt{\alpha}}{4\sqrt{2}}
+\frac{\sqrt{\alpha}}{2\sqrt{2}} \tan^{-1}\left[
\frac{F-4\alpha}{2\sqrt{2\alpha (F-2\alpha)}} \right] + \frac{1}{2}
\sqrt{F-2\alpha},
\end{eqnarray}
Clearly, for a positive $\alpha$, the central shell doesn't get
trapped, and the untrapped region around the center increases with
increasing $\alpha$, for both homogeneous and inhomogeneous models.
Thus the singularities are always naked as they are formed prior to
the formation of apparent horizon.
\section{Discussion}
In this paper, we have found exact spherically symmetric 5D-LTB
solutions, for $W(r)=1$ case, to GB extended Einstein equations
namely 5D-LTB-EGB, which describes gravitational collapse of
spherically symmetric inhomogeneous dust in a 5D spacetime. While
there may be an AH about this singularity, for $\alpha> 0$, the
singularity always remains visible to any observer as the AH lies
beyond singularity which is actually not in the spacetime. It is
interesting that the coupling constant of the GB terms produces a
change in the location of the AH by the factor $2 \alpha$ is exactly
the same as in the case of 5D null fluid collapse in EGB. The final
fate of gravitational collapse is quite different in the sense that
a massive naked singularity is formed, which is disallowed in
5D-LTB. However, the strength singularity is weaker as compared to
the corresponding 5D-LTB \cite{sjsg}, this may not be a serious
threat to CCC. It is seen that the time for the occurrence of the
central shell focusing singularity for the collapse is increased as
compared to the analogous 5D-LTB case. The reason may be, there is
relatively less mass-energy  collapsing in the 5D-LTB-EGB spacetime
as compared to the 5D-LTB case. In particular, our results in the
limit $\alpha \rightarrow 0$ reduce \emph{vis-$\grave{a}$-vis} to 5D
relativistic case.

\ack One of the authors (SGG) thanks University Grants Commission
(UGC) major research project grant F. NO. 39-459/2010 (SR).

\smallskip

\end{document}